\documentclass{PoS}
\title{Modeling the Saturation of the Bell Instability Using Hybrid Simulations}

\ShortTitle{The Saturation of the Bell Instability}

\author{\speaker{Georgios Zacharegkas}
        \\
        University of Chicago, 5640 S Ellis Ave., Chicago, IL 60637 (USA)\\
        E-mail: \email{gzacharegkas@uchicago.edu}}

\author{Damiano Caprioli\\
        University of Chicago, 5640 S Ellis Ave., Chicago, IL 60637 (USA)\\
        E-mail: \email{caprioli@uchicago.edu}}
        
\author{Colby Haggerty\\
        University of Chicago, 5640 S Ellis Ave., Chicago, IL 60637 (USA)\\
        E-mail: \email{chaggerty@uchicago.edu}}

\abstract{The nonresonant cosmic ray instability, predicted by Bell (2004), is thought to play an important role in the acceleration and confinement of cosmic rays (CRs) close to supernova remnants. Despite its importance, the exact mechanism responsible for the saturation of the instability has not been determined, and there is no first-principle prediction for the amplitude of the saturated magnetic field. 
Using a survey of self-consistent kinetic hybrid simulations (with kinetic ions and fluid electrons), we study the saturation of the non-resonant streaming instability as a function of the parameters of both the thermal background plasma and the CR population. The strength of the saturated magnetic field has important implications for both CR acceleration in supernova remnants and CR diffusion in the Galaxy.}

\FullConference{36th International Cosmic Ray Conference\\
                24 July - 1 August 2019\\
                Madison, Wisconsin}

\usepackage[dvipsnames]{xcolor}
\usepackage{graphicx}
\usepackage{amsmath,amssymb}
\usepackage{subfigure}
\usepackage{float}

\begin{document}

\section{Introduction}\label{sec:intro}
High energy cosmic rays (CRs), while relatively few by number, make up a significant fraction of the energy budget of the interstellar medium. Collisionless shock waves associated with supernova remnants are believed to be the primary source of galactic CRs (with energies less than $10^{15}$eV), through a diffusive shock acceleration (DSA) mechanism \cite{bell78a}. For efficient CR acceleration through DSA, CRs must be confined close to the shock, which requires the presence of very strong, turbulent magnetic fields. Strong magnetic turbulence and CR acceleration are thus closely related and the study of the excitation, growth and saturation of such strong magnetic fields is crucial to explain the origin of high-energy CRs.

\cite{winse84} and \cite{bell04} found that in systems with a sufficiently strong CR current, modes with wavelengths significantly smaller than the CR gyroradius could be excited; these modes are different from those driven by the resonant cosmic-ray instability, which is caused by CRs in gyroresonance with Alfv\'{e}nic modes \cite{kulsrud69}. This \emph{nonresonant cosmic-ray instability}, which we refer to as the "Bell instability" in this work, has important implications for DSA at SNRs. Specifically, the highest energy CRs produced by DSA travel upstream along magnetic field lines and drive this instability, which is crucial in order to confine CRs near the shock \cite{caprioli+14b}. 

The dispersion relation of the Bell instability has been calculated using both MHD \cite{bell04} and kinetic treatments \cite{reville06,amato+09} in the context of the linear theory. 
In both cases it is found that for a constant CR current, $\mathbf{J}_{CR}$, the Bell instability grows exponentially with the fastest growing mode, $\lambda_{\max}$, having a growth rate $\gamma_{\max} \propto J_{CR}/B_0$, where $B_0$ is the magnitude of the background ambient magnetic field. In their MHD simulations, \cite{bell04} and \cite{zirakashvili08} found a large amplification factor for the magnetic field, but due to limitations of MHD simulations in modeling large fluctuations they either overestimated the saturation level of the magnetic field or did not observe it at all. Another reason for this is that they did not include the back-reaction on the CR current by the background plasma, which plays an important role in the saturation mechanism of the instability.

Using particle-in-cell (PIC) simulations, \cite{niemiec08} found a much lower level of saturation for the magnetic field fluctuations, but they also found a growth rate for the fastest growing mode which was smaller than what \cite{bell04} had predicted, putting into question the existence of the Bell instability beyond the MHD limit. However, further PIC simulation by \cite{riquelme09} showed that for the exceedingly strong currents adopted in reference \cite{niemiec08}, a transverse, filamentary mode can grow faster than the Bell instability. Nevertheless, \cite{riquelme09} found that for typical CR currents the Bell instability grows as expected and saturates to levels of $\delta B/B_0 \gtrsim 10$. The saturation was found to be caused by the background plasma being accelerated in the direction of the CR drift velocity, which reduces the current $\mathbf{J}_{CR}$ that drives the instability. Similar results have been found in PIC simulations performed by other groups as well \cite{stroman09,ohira09}.

In \cite{gargate10} hybrid (with kinetic ions and fluid electrons) PIC simulations were used to follow the instability on longer time-scales, well into the non-linear regime where the saturation takes place. 
They deduced results for the saturation mechanism similar to \cite{riquelme09}, namely that the saturation occurs due to the deceleration of CRs and simultaneous acceleration of the background plasma, which reduces the CR current. This deceleration of CRs was also seen in the earlier work of \cite{lucek00}.

\section{Theoretical background}\label{sec:theory}
We consider a population of streaming CRs, with number density $n_{CR}$, that carry an electric current density $\mathbf{J}_{CR} = e n_{CR} \mathbf{v}_{CR}$, where $e$ is the charge of a CR and $\mathbf{v}_{CR}$ is the drift velocity of the CRs. In the case where the CRs travel parallel to a background magnetic field $\mathbf{B}_0$ through a plasma of number density $n_g$, they can drive an instability that exponentially amplifies perpendicular magnetic field perturbations. 
The analytical dispersion relation for the Bell instability \cite{bell04} predicts a maximum growth rate of
\begin{equation}\label{eqth:gamma_max Bell}
	\gamma_{\max} = k_{\max} v_A = \frac{1}{2} \left( \frac{n_{CR}}{n_g} \right) \left( \frac{v_{CR}}{v_A} \right)\Omega_{ci} \; ,
\end{equation}
of the fastest growing mode, which reads
\begin{equation}\label{eqth:k_max Bell}
	k_{\max} = \frac{4\pi}{c} \frac{J_{CR}}{B_0} = \frac{1}{2} \left( \frac{n_{CR}}{n_g} \right) \left( \frac{v_{CR}}{v_A} \right) d_i^{-1} \; ,
\end{equation}
where $v_A = B_0/\sqrt{4\pi m n_g}$ is the Alfv\'{e}n speed based on the background plasma density and magnetic field, $\Omega_{ci}=eB_0/mc$ is the ion gyrofrequency, $m$ is the ion mass, $c$ is the speed of light, and $d_i = v_A/\Omega$ is the ion inertial length (equivalent to $c/\omega_{pi}$, where $\omega_{pi}$ is the ion plasma frequency). 

This prediction assumes that the thermal effects on the dispersion relation are negligible, i.e., the background plasma's gyroradii must be much smaller than the wavelength of the amplified magnetic fluctuations, $k_{\max}r_g \ll 1$. However, as $k_{\max} r_g$ increases, thermal effects reduce the growth rate of the fastest growing mode and lead to a modified instability referred to as the WICE instability, discussed in \cite{zweibel+10}. Although these instabilities are important for generating magnetic fields in astrophysical systems, it has not been well established what causes their saturation or what is the strength of the saturated magnetic field. Bellow we discuss proposed theories for the saturation of the Bell and WICE instabilities.

\subsection{Energy equipartition hypothesis}\label{subsec:EnergyEquipartition}
Based on this hypothesis, the saturation of the instability occurs once the magnetic energy density reaches equipartition with the CR energy density \cite{bell04}. Quantitatively, this implies that at saturation the kinetic energy density of the CRs, $\mathcal{E}_{CR} = n_{CR} m c^2(\gamma_{CR} - 1) \approx n_{CR}v_{CR}p_{CR}$, where $p_{CR}$ is the momentum corresponding to the drift velocity of the CRs and $m$ the mass of a CR particle, becomes comparable to the magnetic energy density, $\mathcal{E}_{B} = \delta B^2/8\pi$, assuming that $\delta B$ (the perpendicular amplified magnetic field) has reached highly non-linear levels. Here, $\gamma_{CR}$ denotes the average Lorentz factor of the CRs.

\subsection{Resonance Hypothesis}\label{subsec:ResonanceHypothesis}
An alternative scenario for saturation proposes that the magnetic field stops growing when the CR current is sufficiently reduced by resonant pitch-angle scattering of the CRs \cite{caprioli+14b}. This should occur when $k_{\max} r_{CR}' \approx 1$, where $r_{CR}'$ is the CR gyroradius in the amplified magnetic field, which is expected to be smaller than the original gyroradius by more than a factor of unity. Note also that as $\delta B \gtrsim B_0$, the instability enters a non-linear growth stage and $k_{\max}$ may depend on $\delta B/B_0$ \cite{riquelme09,caprioli+14b}.

\subsection{Movement of the Background Gas}\label{subsec:GasMovement}
Kinetic simulations have shown that the acceleration of the background plasma in the direction of the CR drift velocity during the non-linear stage should play a fundamental role in the saturation of the Bell instability \cite{riquelme09,gargate10}. When the magnetic field reaches non-linear levels, the CRs are effectively scattered by the field fluctuations. Due to conservation of momentum, this leads to transfer of momentum to the magnetic field and ultimately to the background gas, thus accelerating the gas in the direction of $\mathbf{v}_{CR}$. As a result, the Bell instability likely transitions to the resonant streaming instability (the weak current regime, \cite{amato+09}) and soon saturates.

\section{Hybrid Simulations}\label{sec:sim description}
To study the saturation of the instabilities discussed above we preform kinetic hybrid simulations with {\it dHybridR}, in which the ions are treated as macro-particles governed by the relativistic Lorentz force law and the electrons are a massless, charge neutralizing fluid \cite{haggerty+19a}. In our simulations the background magnetic field is oriented along the $x$-direction. Henceforth, we will refer to this as the {\it parallel} direction denoted by the symbol "$||$", while the direction perpendicular, i.e., the $y\ \&\ z$ direction will be {\it perpendicular}, denoted by "$\perp$". We have run 18 2D simulations to study how the different CR and background plasma parameters affect the saturation. In each system CRs were injected as a beam into the simulation box from the left boundary. The CR distribution is defined by a drift velocity in the $x$ direction, of magnitude $v_{CR}=p_{CR}/m\gamma_{CR}$, plus an isotropic momentum of modulus $p_{iso}$. In this work we only consider very cold CR distributions with $p_{iso}=m v_A$ and relativistic CRs with $p_{CR}\geq mc$ (see Table \ref{tab:sims}). The CRs, therefore, have a current density $\mathbf{J}_{CR}= e n_{CR} \mathbf{v}_{CR}$ and pressure given by the tensor
\begin{align}\label{eqsim:pressure tensor}
	P_{ij,CR}(t) = \int  v_i p_j f_{CR}(\mathbf{p},t) d^3p \;  ,
\end{align}
where $f_{CR}(\mathbf{p},t)$ is the CR distribution function in momentum space. In what follows we will denote the parallel component of the pressure simply by $P$ and the perpendicular component by $Q$. Moreover, the index ''0'' will indicate initial values in any quantity it might appear.

\begin{table}[ht]
    \centering
    \begin{tabular}{|c|c|c|}
        \hline
        Simulation & $p_{CR}/mv_A$ & $n_{CR}/n_g$\\
        \hline
        \hline
        A & $100$ & $[10^{-4},10^{-1}]$\\
        \hline
        B & $1000$ & $[10^{-4},10^{-1}]$\\
        \hline
        C & $1000$ & $5 \times 10^{-3}$\\
        \hline
        D & $1000$ & $5 \times 10^{-2}$\\
        \hline
    \end{tabular}
    \caption{Simulations used in this work. The column $p_{CR}/mv_A$ corresponds to the parallel CR momentum component, while the next column shows the CR number density $n_{CR}$ normalized by the gas number density $n_g$. Note that for simulations A and B $n_{CR}$ lists the range of the CR number density used in a series of 16 runs in total. The speed of light in units of the Alfv\'en speed is $c=100 \; v_A$, the dimensions of the 2D box of our simulations are $L_x = 1000 \; d_i$ and $L_y = 1000 \; d_i$, $\Delta x = 0.5 \; d_i$ is the size of one cell, while $\Delta t = 2.5 \times 10^{-4} \; \Omega_{ci}^{-1}$ represents the time step of integration. For all simulations we considered CRs with isotropic momentum $p_{\rm iso} = mv_A$.}
    \label{tab:sims}
\end{table}

Our simulations are 2.5D (2D in real space with all 3 components of the velocity and fields retained) with periodic boundary conditions for the fields, thermal particles and for the perpendicular motions of the CRs, and open boundary conditions for the CRs along the $\pm x$ directions. This configuration is physically comparable to SNR environments in which energetic CRs are being continuously injected by the shock wave, a setup previously neglected in simulations campaigns. We have tested various simulations sizes to verify that the saturation of $\delta B$ is not affected, and we have found that convergence of the saturated field occurs when the parallel and perpendicular lengths are at least one CR gyroradius, defined based on the initial magnetic field.

\section{Results}\label{subsec:SimResults}
To understand the saturation of the non-resonant instability we have preformed simulations that cover a wide range of our parameter space (partially shown in Table \ref{tab:sims}). 
We have identified two distinct regions of the parameter space that give rise to the Bell and the WICE instability; 
the transition from the Bell to the WICE regime depends on the CR current density (which sets the fastest growing mode) and it roughly corresponds to $J_{CR}/e n_g v_A \approx 1$. 
Specifically, when this value is less than 1 the standard Bell instability is triggered, while $J_{CR}/en_gv_A \gtrsim 1$ corresponds to the WICE regime. 
This suggests that for large CR currents, the growth rate of the instability should be suppressed relative to the Bell prediction. 

\begin{figure}
	\centering
	\includegraphics[width=3.5in]{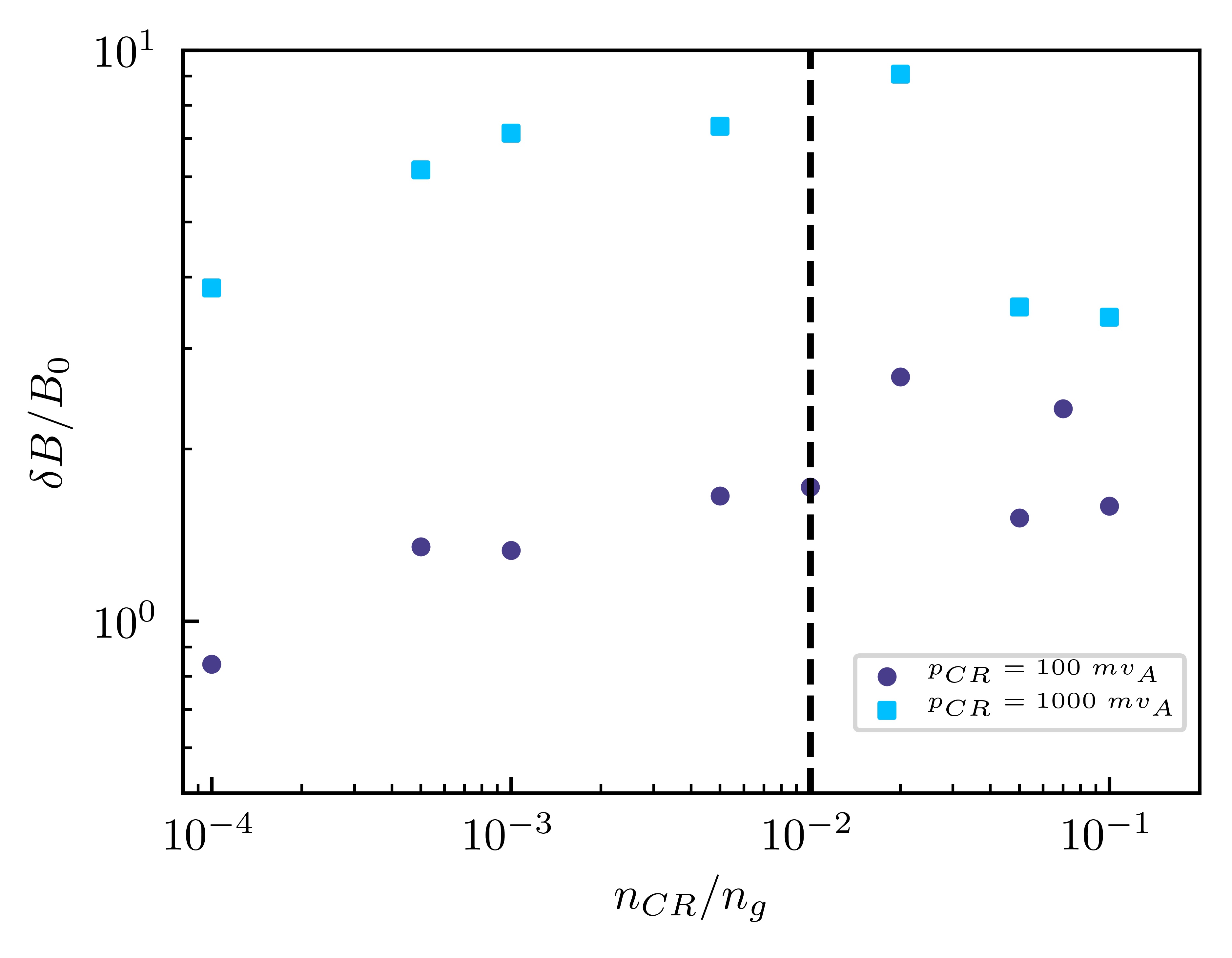}
	\caption{Average magnetic field at saturation as a function of CR number density for simulations A and B in Table \ref{tab:sims}. For these runs we have fixed the parallel CR momentum to $p_{CR} = 100 \; mv_A$ (blue circles) and $p_{CR} = 1000 \; mv_A$ (cyan squares). The vertical dashed line corresponds to $J_{CR} = 1 \; en_gv_A$ where the transition from the Bell to the WICE regime is expected to occur.}
	\label{fig:Bperp_vs_ncr}
\end{figure}

Figure \ref{fig:Bperp_vs_ncr} shows the value of the amplified magnetic field at saturation as a function of the injected CR number density for the two sets of simulation A and B in Table \ref{tab:sims}. 
Note that the CR density in these simulations varies from $n_{CR}/n_g = 10^{-4}$ to $n_{CR}/n_g = 10^{-1}$ and thus the CR current, $J_{CR} \approx e n_{CR} p_{CR}/\gamma_{CR}$, varies from $\sim 10^{-2}$ to $\sim 10$ in units of $en_gv_A$. The saturated magnetic field increases with $n_{CR}$ until $J_{CR} \approx 1 \; en_gv_A$, above which the instability is in the WICE regime and the magnetic field plateaus or even decreases as $n_{CR}$ increases. In addition to a transition in the saturation behaviour, there is also a clear difference in the magnitude of the saturated field between the $p_{CR}/mv_A = 100$ and $1000$ cases (or equivalantly $\gamma_{CR}=\sqrt{2} \text{ and } 10$ cases). It is clear that the larger $p_{CR}/mv_A$ is the larger the saturated magnetic field.

\begin{figure}
	\centering
	\includegraphics[width=6.in]{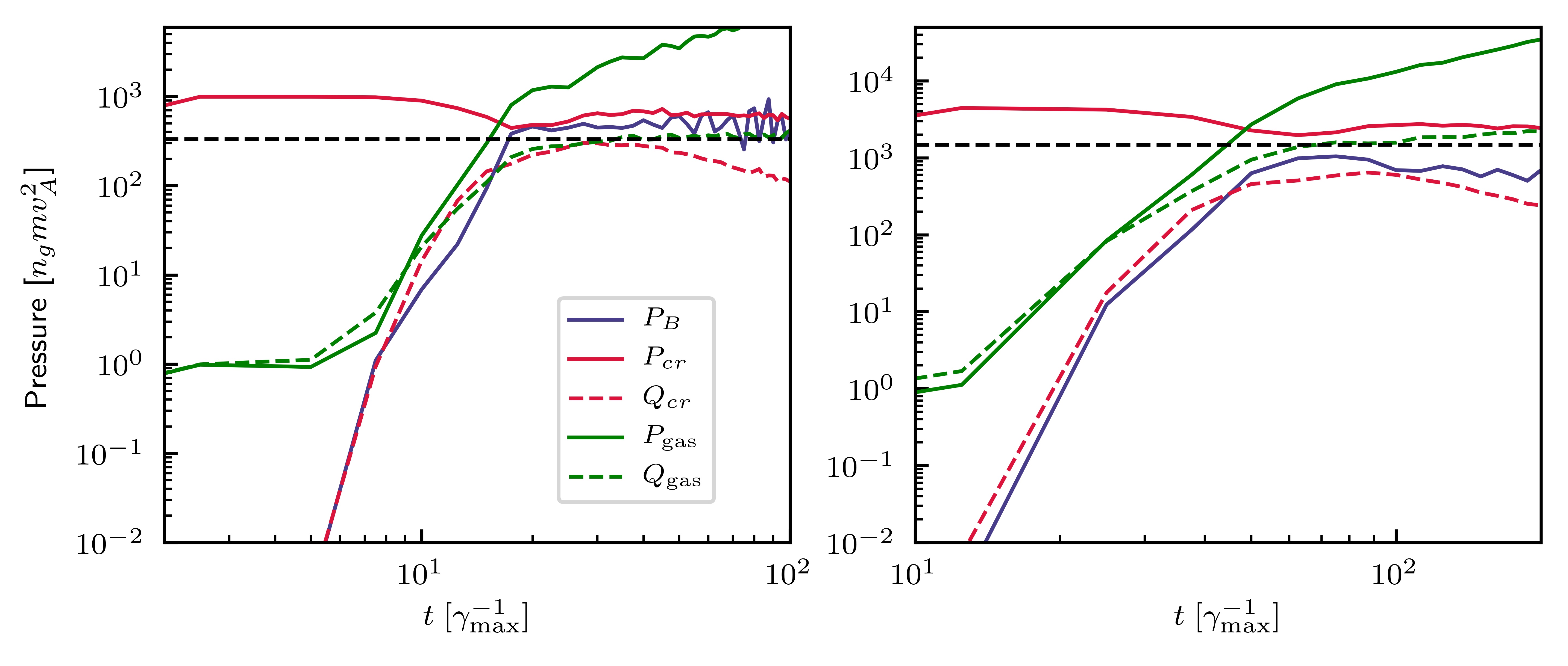}
	\caption{Time evolution of the different pressure components averaged over the simulation domain for simulation C (left) and D (right). 
	Red/green indicate CR/gas pressures, and solid/dashed lines correspond to parallel  and perpendicular pressures (see Eq.~\eqref{eqsim:pressure tensor}); $P_B = \delta B^2/8\pi$ is shown in blue. 
	In the left panel $J_{CR} = 0.5e n_gv_A$ and in the right panel $J_{CR} = 5en_gv_A$. The horizontal dashed line indicates the $P_{CR,0}/3$ level.}
	\label{fig:pressures}
\end{figure}

Figure \ref{fig:pressures} shows how the thermal, CR and (perpendicular, i.e., self-generated) magnetic pressure evolve in time. The left plot corresponds to a simulation in the standard Bell regime, with $J_{CR} = 0.5 \; en_gv_A$ (simulation C in Table \ref{tab:sims}). The magnetic field initially grows exponentially until $P_B \sim mn_gv_A^2$, at  $t\sim 7 \; \gamma_{\rm max}^{-1}$, which marks the beginning of the ``secular'' stage of the instability, when $\delta B$ keeps growing but with a slower growth rate. During this stage there is a rapid increase in perpendicular gas pressures, $Q_{\rm gas}$, as well as in the parallel pressure which is associated with the acceleration of the background gas. Such an acceleration is the result of the CRs transferring momentum to the gas by scattering off of the magnetic field fluctuations. This can be seen in the drop of the parallel CR pressure, $P_{CR}$, at $t \sim 10 \; \gamma_{\max}^{-1}$. The secular phase continues until $Q_{\rm gas}$ reaches the level of $\sim P_{CR,0}/3$, at about $t\sim 20 \gamma_{\rm max}^{-1}$. The end of this phase predictably coincides with the saturation of the magnetic field. In this simulation it is clear that the system reaches a state of pressure equilibrium where $Q_{\rm gas} \sim P_B \sim P_{CR,0}/3$. 

The right plot in figure \ref{fig:pressures} shows the same quantities, but for a simulation with $J_{CR} \approx 5 \; en_gv_A$ (simulation D from Table \ref{tab:sims}). The initial exponential growth phase is similar to the previous, weak-current case. Also similarly, the magnetic field saturates when $Q_{\rm gas} \sim P_{CR,0}/3$; however, in this case the magnetic pressure is not in equipartition with the other two pressures. This may  be attributed either to the fact that in this (WICE) regime the growth rate of the magnetic field is suppressed by thermal effects, or that some magnetic damping mechanism becomes important.

\begin{figure}
	\centering
	\begin{minipage}{0.45\textwidth}
		\centering
		\includegraphics[width=1.\textwidth]{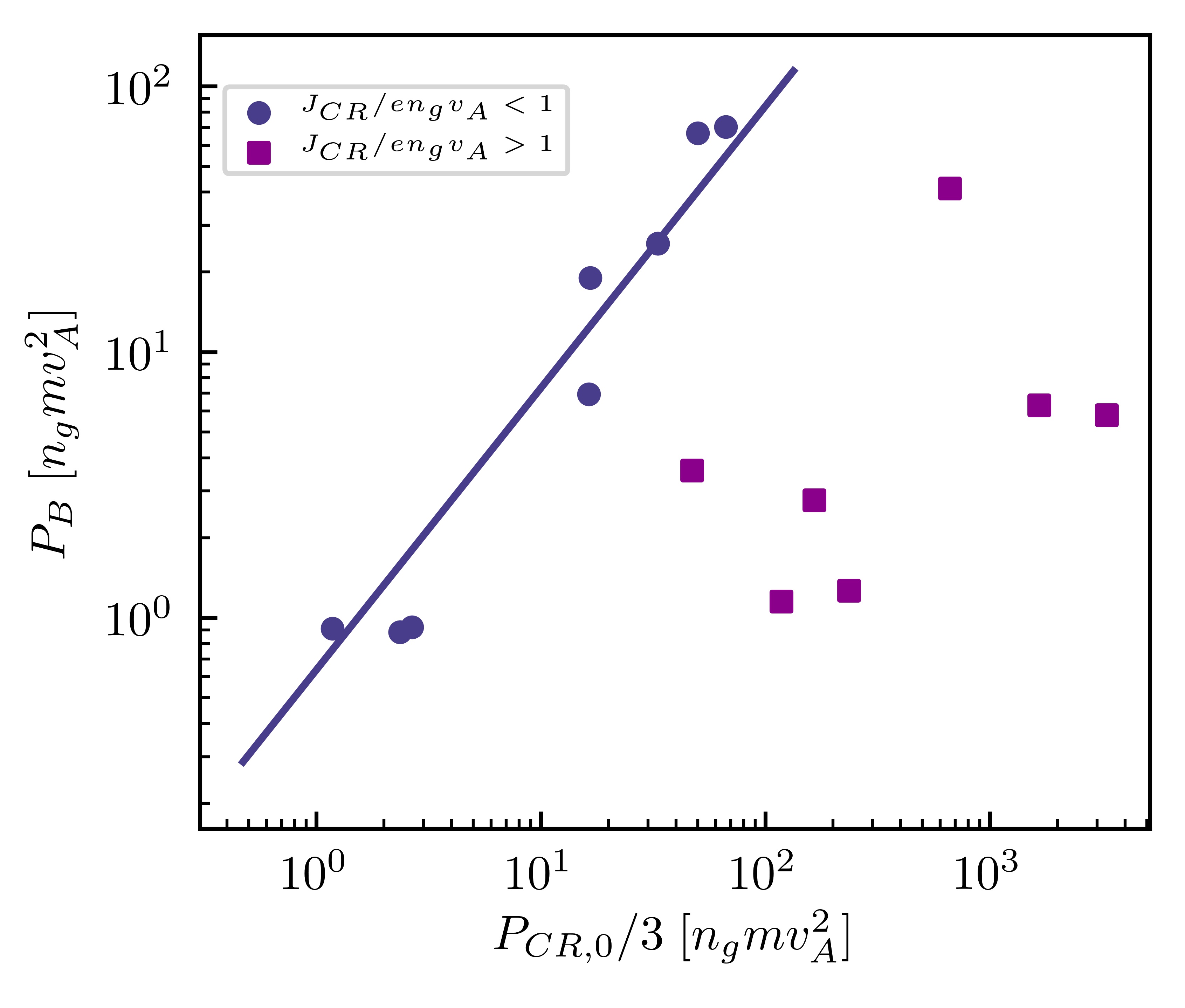}
		%			\caption{first figure}
	\end{minipage}\hfill
	\begin{minipage}{0.45\textwidth}
		\centering
		\includegraphics[width=1.\textwidth]{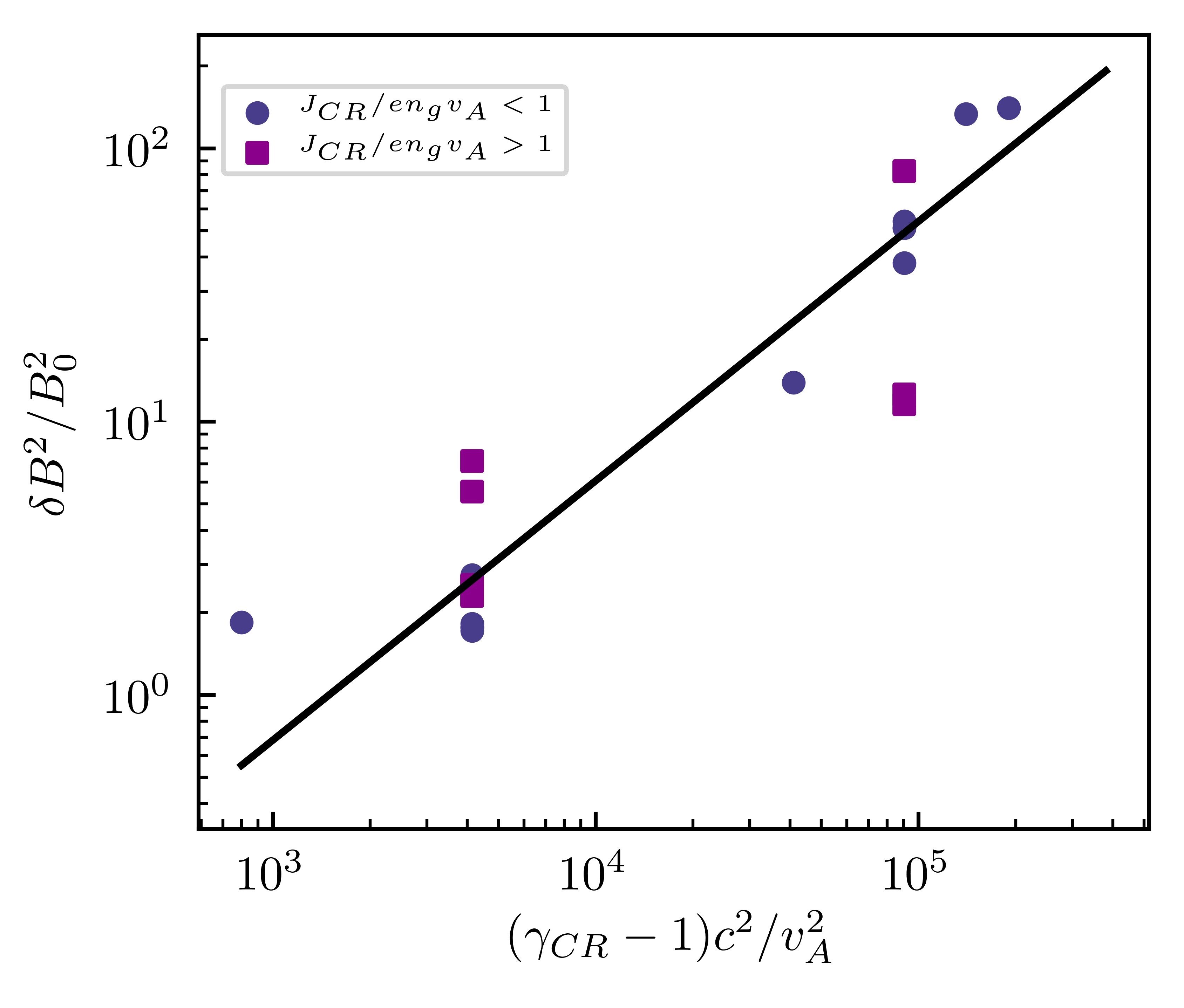}
		%			\caption{second figure}
	\end{minipage}
	\caption{\label{fig:saturation_plot} {\it Left:} Saturated magnetic pressure as a function of $P_{CR,0}/3$ for the simulations with weak (blue circles) and strong (purple squares) CR currents. The slope of the best-fit line in this log-log plot for the low-CR current points is $\sim 1.061$ and the intercept is $\sim 0.64$, showing that the two are linearly proportional and differ only by a factor of unity. Note that simulations in the WICE regime fall well below such a correlation. {\it Right:} Final magnetic field amplification for weak and strong CR currents as a function of the kinetic energy per CR particle, normalized by $mv_A^2$.}
\end{figure}

The saturated magnetic field for various simulations is shown in Figure \ref{fig:saturation_plot}. The complete list of runs used in these plots is too extensive to be included in this conference proceeding, and it will be presented in a forthcoming work. The left panel shows the final magnetic field pressure as a function of $1/3$ of the initial CR pressure, calculated as $P_{CR,0} \approx \gamma_{CR} n_{CR} m v_{CR}^2$, normalized in terms of $n_g m v_A^2$. Markers distinguish simulations with weak ($J_{CR} < 1$, blue circles) and strong ($J_{CR} > 1$, purple squares) CR currents normalized to $en_g v_A$. Consistent with the previous discussion, for the simulations in the standard Bell regime the saturated magnetic pressure is proportional to the saturated CR pressure. The best fitting is indeed linear, with a normalization factor of $\sim 2$ below the pressure equipartition. For the strong CR current simulations ($J_{CR} > 1$), the points fall quite below equipartition and $P_B$ is not correlated with $P_{CR}$. 

The right panel of Figure \ref{fig:saturation_plot} shows the final amplification factor (proportional to the final magnetic pressure and energy density) as a function of the initial CR kinetic energy \emph{per particle}, $ (\gamma_{CR} -1) mc^2$, measured in units of $mv_A^2$. If we introduce the CR energy density in code units as $\epsilon_{CR}$ the quantity on the horizontal axis of Figure \ref{fig:saturation_plot} is simply $\epsilon_{CR}/n_{CR}$. There is a clear linear correlation between the final saturated magnetic energy density and the kinetic energy of the beam, for both the weak and strong CR current regimes. While this scaling is rather clear, the physical explanation for this relationship is not trivial and will be discussed in a forthcoming publication. 

\section{Conclusions}
We used hybrid simulations preformed with {\it dHybridR} to study the growth and saturation of the non resonant (Bell) streaming instability. We identify a transition in the behaviour of the saturated magnetic fields based on the strength of the CR current density relative to $en_{g}v_A$. For weak CR currents ($J_{CR}/en_gv_A < 1$), the instability saturates when the magnetic and CR pressures become comparable, while for strong CR currents ($J_{CR}/en_gv_A > 1$, WICE regime), the saturated magnetic field is not well correlated with the CR pressure. 

In general, we find that at saturation the gas pressure is almost in equipartition with the CR pressure for both regimes. Finally, we outline a correlation between the saturated magnetic energy density and the initial CR energy per particle, a result that does not have a trivial theoretical counterpart and that will be further examined in a forthcoming publication.

%\clearpage
\bibliographystyle{JHEP}
%\bibliography{Total}
\bibliography{Bell_bib,Total}

\end{document}